\documentclass[prd,nofootinbib,twocolumn]{revtex4}
\usepackage{amsmath}
\usepackage{amssymb}
\usepackage{graphicx}
\usepackage[usenames, dvipsnames]{color}

\newcommand{\sgn}{\operatorname{sgn}}
\usepackage{graphicx}

\begin{document}

\title{Issues in the comparison of particle perturbations and numerical relativity for binary black hole mergers}

\author{Richard H.~Price} 
\affiliation{Department of Physics, MIT, 77 Massachusetts Ave., Cambridge, MA 02139}
\author{Gaurav Khanna} 
\affiliation{Department of Physics, East Hall, University of Rhode Island, Kingston, RI 02881}
\affiliation{Department of Physics and Center for Scientific Computing \& Data Research, University of Massachusetts, Dartmouth, MA 02747}

\begin{abstract}
Recent work on improved	efficiency of calculations for extreme mass
ratio inspirals has produced the useful byproduct of comparisons of
inspirals of {\it comparable} mass by particle perturbation (PP) methods and
by numerical relativty (NR). Here we point out:  (1) In choosing the rescaling
of the masses, consideration must be given to the differences in the
PP and NR methods even in the earliest, least nonlinear regime; in particular 
barycenter effects must be addressed. (2) Care must be given to the comparison 
of the nonspinning remnant in PP and the rapidly spinning remnant in NR.
\end{abstract}

\maketitle

\section{Introduction}\label{sec:Intro}
Next generation interferometric gravitational wave (GW) detectors will be
sensitive to signals from intermediate/extreme mass-ratio inspirals (I/EMRIs)
of binary black holes. For such sources a numerical relativity (NR)
solution of the full nonlinear Einstein equations may prove to be
challenging and may not be needed. Other modeling methods for I/EMRIs include 
point-particle black hole perturbation theory~\cite{PP,Harms}, the gravitational 
self-force~\cite{SF1,SF2,SF3,SF4,SF5,SF6} and effective-one-body~\cite{EOB1,EOB2,EOB3,EOB4,EOB5,EOB6,EOB7}. 
High accuracy and improved insights are likely to come from particle perturbation 
(PP) methods. Even these methods, however, are computationally expensive. To reduce 
the computational cost of NR and I/EMRI calculations, 
data-driven ``surrogate'' models have been developed~\cite{Surr1, Surr2,Surr3}. 
Machine learning based surrogate models presented in Refs.~\cite{Surr2,Surr3} 
are ``trained'' using PP waveforms calibrated with NR results for various values 
of the ratio $q$ of $M_1$, the more massive hole, to the smaller hole $M_2$.

Every method is limited in the value of $q\equiv
M_1/M_2$ for which it is appropriate. 
For $q$ much greater than approximately $10$, NR cannot easily handle the great
disparity in grid size around the two black hole regions.  PP on the
other hand, treats $M_2$ as a perturbation on the spacetime of
$M_1$. It is therefore appropriate for the EMRI limit and may yield 
astrophysically reasonable results for $q$ much greater than around $10$. 
But, despite low expectations of accuracy, there is nothing that prevents
PP codes from running for smaller values of $q$, i.e., for comparable
mass holes, and from comparing the results to those of NR. It has been
somewhat surprising that the results of these PP explorations, when
simply rescaled, agree rather well with the NR results~\cite{Surr2,Surr3}.

Because black hole processes scale in black hole mass, confusion lurks as a possibility 
in interpreting both NR and PP data.  The numbers reported in both cases, what might be 
called ``raw'' data, have to be multiplied by a mass to produce dimensional data. The 
relevant dimensional data in both cases are the periods of the GW oscillations and the 
product of the (dimensionless) GW strain $h$ with the (dimensional) radius $r$ at which 
the GW data are extracted.

A correct interpretation of conversion from raw to dimensional numbers is that in the NR 
case, the raw numbers are to be multiplied by the total mass of the spacetime $M_{\rm total}=M_1+M_2$, 
while the PP results are to multiplied by the larger mass $M_1$. 

If we want to compare physically equivalent configurations being computed in the two methods, 
then we must arrange for them to refer to the same value of $M_1$. If we were to compare the ``raw'' 
data we would be comparing PP and NR data for different $M_1$. To correct for this, the raw data 
for PP must be  reduced to a scaling corresponding to the $M_1$ of the NR computations. This means 
reducing the PP raw data by a factor $M_1/M_{\rm total}=1/(1+1/q)$ This is what was done in the 
surrogate papers, and what we, like those authors, will call the naive rescaling.

For our primary example, $q=3$, this naive reduction factor is $0.75$.
We compared the NR\footnote{It should be noted that NR can't 
produce long wave trains, therefore we use a hybridized surrogate model that
combines late stage NR data with the effective-one-body (EOB) model to 
generate long-duration inspiral waveforms. We will refer to the hybrid models simply 
as NR, since the issues we discuss (alignment, barycenter, QNR) apply in the 
same way to the hybrid results as to NR results.}~\cite{hybrid} and PP raw 
data~\cite{PP} for $q=3$ from very early Newtonian orbits, with a separation 
on the order of $3GM_{\rm total}/c^2$, to very late quasinormal ringing -- 
for the entire range of cycles we find a NR/PP ratio of gravitational wave periods
$0.79\pm0.2$. 

\begin{figure}[h!]
\caption{Comparison of NR results (dashed-black curve) for gravitational wave amplitude, and ``naively'' 
rescaled PP results (solid-red curve) for $q=3$. The rescaling is a reduction of both the time and the 
gravitational wave strength $rh_{+}$ by the factor of $1/(1+1/q)=0.75$. The curves are aligned so that 
the peaks of both the NR and PP runs agree.
}
\includegraphics[width=.25\textwidth]{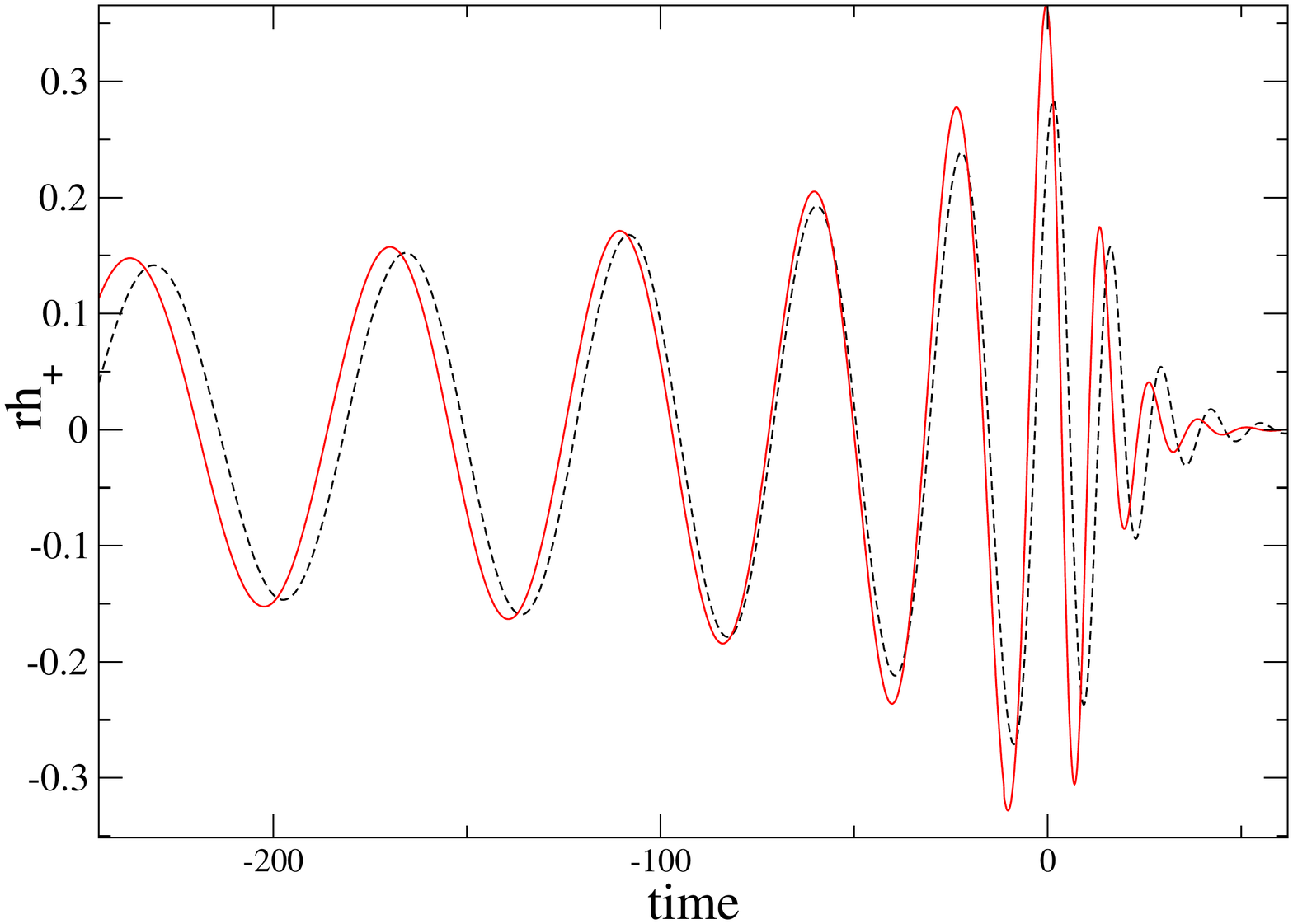} \hspace{-.28in}
\includegraphics[width=.25\textwidth]{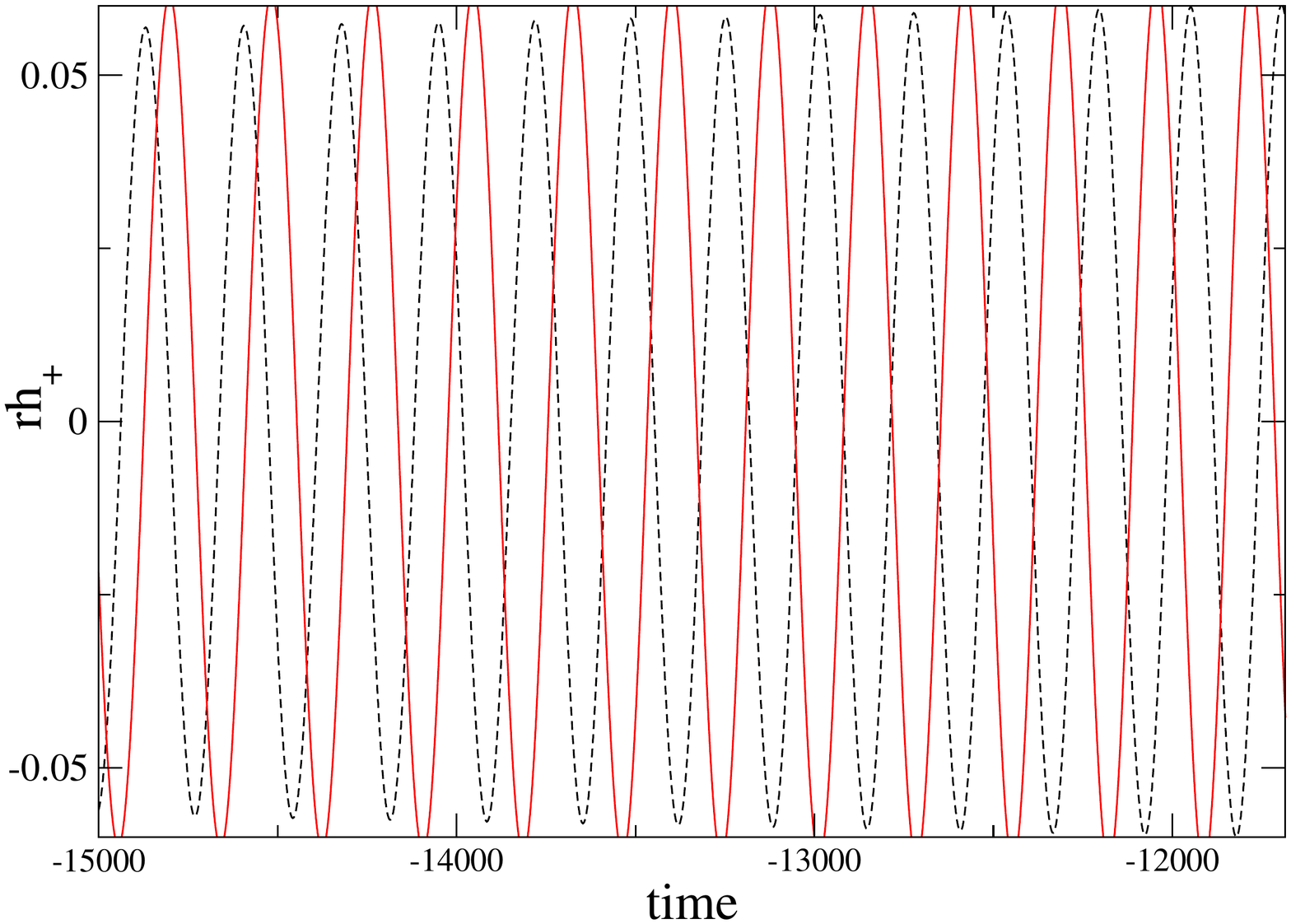} \label{naive}
\end{figure}

For $q=3$ we show the result of naive scaling in Fig.~\ref{naive}. The dashed-black curves are the 
NR results. The solid red curves are the PP results rescaled with reduction of values on both axes by the 
naive scaling  $1/(1+1/q)=0.75$. The curves are aligned so that the peaks of both the NR and PP runs 
agree. It is clearly seen that although the amplitudes are in rough agreement, the phases are not. 


The purpose of this paper is to discuss the considerations for a 
proper NR/PP comparison. This is a needed first step if the physics
of nonlinearities (beyond the {\em adiabatic} radiative corrections) 
is to be extracted from the comparison.

The considerations will be presented in three separate sections: the
problem of aligning waveforms in Sec.~\ref{sec:align}; the inclusion
of barycenter effects in Sec.~\ref{sec:barycenter}; the comparison of
quasinormal ringing and considerations of the spin of the remnant, in
Sec.~\ref{sec:QNR}; and comments on the relation of our results with 
those from optimization~\cite{Surr3} in Sec.~\ref{sec:Comp}. Some 
mention of possible future directions is given in Sec.~\ref{sec:Conclusions}.

\section{Overall: Alignment}\label{sec:align}
The gravitational waveforms from NR and PP are a list of raw times and
the corresponding raw values for the gravitational wave amplitudes
$r\,h_{+}$.\,\footnote{We have found nothing different about what can be 
extracted from $h_{\times}$, the other polarization, so for simplicity we 
limit discussion to $h_{+}$} Since some rescaling, whether naive or sophisticated, 
will be needed, it should be clear that simply comparing the waveforms at 
equivalent raw times does not amount to comparison at equivalent physical 
configurations. (The small range $0.79\pm0.2$ of the NR/PP period ratio, 
therefore, offers little physical insight.)  
Previous attempts to compute the rescaling via an optimization procedure~\cite{Surr2,Surr3}
while it led to an unexpected  agreement with NR, were also not physically 
well motivated. 

\begin{figure}[h!]
\caption{Aligned rescaled data for $q=3$. NR data is the solid-red curve and the 
  rescaled PP is depicted by the dashed-black one. Rescaling is done by reducing 
  both the period and GW amplitude by the ``naive'' scaling factor $(1+1/q)$, 
  due to the different total masses in NR and PP results for the same $M_1$. The 
  PP periods are then decreased by an additional factor $(1+1/q)^{1/2}$ due to barycenter 
  effects (explained in Sec.~\ref{sec:barycenter}) that have been omitted from previous PP 
  work. Axis are in units of $M_{\rm total}$. 
}\label{align2800}
\includegraphics[width=\columnwidth]{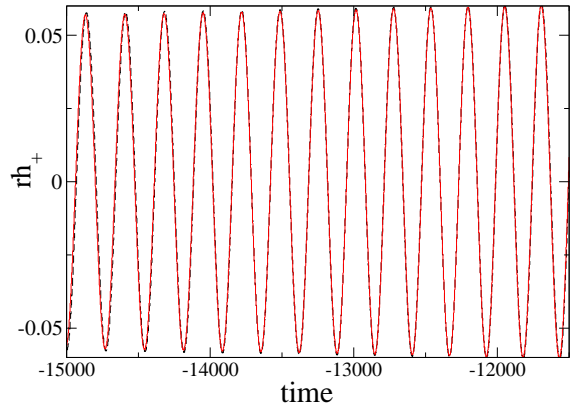}
\end{figure}

What then is the right way to make the comparison? For the earliest GW
cycles the orbits creating the waveforms are very nearly
Keplerian. Subsequent differences between the PP and NR waveforms can
then be understood as signs of nonlinear effects beyond the adiabatic radiative correction. 
Figure~\ref{align2800} shows the results when this is
done with the $q=3$ data for NR and PP. This alignment requires the
use of the rescaling, explained in the next section. This is completely 
``theoretical'' rescaling based on Keplerian orbits, with no optimization.  
When the theoretical rescaling is applied to the GW periods in the PP data, 
the NR waveforms are searched for the same period. In the case illustrated
in Fig.~\ref{align2800}, the earliest period used in the PP data was
around the raw time of $-28000$, and the raw period\footnote{Found by subtracting 
the raw times between two consecutive peaks or troughs.} is approximately 
$430$. The corresponding theoretically rescaled period is $280$. 
A search was then made in the NR data for this period, and was found at time around 
$-15000$. The rescaled PP data was then moved so that the GW 
oscillations overlap. {\em The agreement of the rescaled  GW amplitude in the PP data and 
the GW amplitude in the NR data confirms the validity of the procedure}.

While some effort was made in order to test the robustness of this procedure, 
it should be noted that for very long waveforms that involve many inspiral cycles 
with very slowly varying periods, it may be difficult in practice to perform an 
alignment robustly based on the approach taken above. 

\section{Early cycles: Barycenter}\label{sec:barycenter}
In the early cycles both NR and PP computations have negligible
nonlinear effects, but they are not based on the same models. In
particular, PP treats the binary dynamics as if the coordinate origin,
the center of $M_1$, were an inertial point. This of course is not 
physically correct. It is the barycenter of the two black holes that 
is the inertial point.  With $R_1,R_2$ representing, respectively, 
the distance from the barycenter to $M_1,M_2$ a very straightforward 
calculation gives the angular velocity about the barycenter to be
\begin{equation}
  \Omega_{\rm bar}=\sqrt{\frac{M_1}{R_2\, s^2}\;}
\end{equation}
where $s\equiv R_1+R_2$. For the PP analysis the angular velocity around 
the coordinate origin is
\begin{equation}
  \Omega_{\rm PP}=\sqrt{\frac{M_1}{s^3}\;}\,.
\end{equation}
From these we find that the ratio of the period about the barycenter to 
that about $M_1$ is
\begin{equation}\label{Tratio}
  \frac{T_{\rm bar}}{T_{\rm PP}}=\sqrt{\frac{R_2}{s}\;}=\sqrt{\frac{1}{1+1/q}\;}.
\end{equation}
The overall rescaling of the PP period is then $(1+1/q)^{-3/2}$, consisting 
of a decrease by a factor $(1+1/q)$, due to the change of units and an
decrease by $(1+1/q)^{1/2}$ from the barycenter omission of the PP method.

We need to ask next how the barycenter omission affects the GW amplitude 
in the PP work. To do this we use the ``slow motion'' approximation~\cite{RevModPhys} 
and compute the ratio of $\ell-$pole moments in the PP and NR approaches:
\begin{equation}\label{Qratio}
  \frac{Q^\ell_{\rm bar}}{Q^\ell_{\rm PP}}=\frac{M_1 R_1^{\ell}+M_2 R_2^{\ell}}{M_2 s^\ell} 
  =\frac{1+1/q^{\ell-1}}{(1+1/q)^\ell}\,.
\end{equation}
The GW amplitude in the $\ell$-th moment is proportional to the $\ell$-th time 
derivative of $Q^\ell$ so, with Eq.~\eqref{Tratio}, we have
\begin{equation}\label{hratio_general_ell}
    \frac{h^{\ell}_{\rm bar}} {h^\ell_{\rm PP}}  = \frac{1+1/q^{\ell-1}}{(1+1/q)^{\ell/2}}\,,
\end{equation}
and the interesting conclusion that for {\em the dominant quadrupole radiation the GW amplitude 
is not affected by the omission of barycenter effects in the PP computation}. The gravitational 
wave amplitude must only be scaled by the naive unit scaling. This has been done in Fig.~\ref{align2800}; 
{\em the PP period has been divided by $(1+1/q)^{3/2}$ and the amplitude has been reduced by $(1+1/q)$}.

It is natural to ask how the agreement shown in Fig.~\ref{align2800}
compares with the agreement for naive scaling.  That is, how does the
NR waveform compare to the PP waveform if the alignment is adjusted
for a best fit to the period?  The brief answer is that the fit is
about equally good at least in one sense: In both cases the NR and PP
waveforms drift $180^\circ$ out of phase after around $35$ GW periods.

A more careful answer involves several considerations. Most important,
over the $\approx35$ periods of alignment drift, both the NR and PP periods
change by around $23$\%. This is a much larger drift of phase than what is
observed in the NR vs. PP comparison.

The comparison of ``fit''	then is	not a good basis for a quick
evaluation of the naive	vs. improved scaling. What {\emph is} a good
basis is a comparison of amplitudes for the best fits. As shown in
Fig.~\ref{align2800}, a best fit to the period, gives excellent agreement
with the amplitude. By comparison, for a best fit in the naive case,
the PP amplitude is larger by 10.7\% than the NR amplitude.

Underlying the match in Fig.~\ref{align2800} is the assumption that Keplerian orbits 
describe the motions in both the PP and NR work and linearized general relativity is adequate 
for the GW amplitude. It is interesting therefore to note that the simple quadrupole formula~\cite{MTW36.20} 
underestimates the GW amplitude by around $10$\%, much larger than would be suspected from Fig.~\ref{align2800}
It should also be noted that for $\ell=3$, the rescaling implied by Eq.~\eqref{hratio_general_ell}, along 
with the naive rescaling, implies a reduction of the PP raw data by a factor of $0.54$, whereas the raw 
data, when aligned, show a ratio of around $0.41$. 


\section{Late cycles: quasinormal ringing}\label{sec:QNR}

At the end of both the NR and PP waveforms there is the
characteristically strongly damped, high frequency oscillations of
quasinormal ringing (QNR). It is noted in Ref.~\cite{Surr2} that a rescaling 
close to the naive rescaling brings the QNR of the PP waveform into agreement with the
QNR of the NR waveform. That is an intriguing result, especially if one appreciates  
that the remnant in the NR work is a rotating Kerr hole, while the PP calculation 
maintains the Schwarzschild form it started with. To underscore this, Table~\ref{tab:QNRscaling} 
shows that this kind of rescaling works well only for $q\gtrsim5$ and for these values 
we view the agreement to be simply due to the fact that the there is little variation 
for $5<q<\infty$ and the remnant QN oscillation frequency is expected to increase 
with decreasing $q$~\cite{BertiCardosoDataFiles}. 
\begin{table}[h]
  \caption{Comparison of the QNR frequency for a remnant and the
    rescaled Schwarzschild QNR frequency.  For each value of $q$ in
    the first column, the second column gives the remnant value of the
    spin parameter $a$ listed in the SXS catalog~\cite{SXScatalog} for 
    the SXS model, given in the third column, a model with negligible initial spins.
    The fourth column gives the real part of the least damped $\ell,m=2,2$ 
    quasinormal frequency, from the tables in Ref.~\onlinecite{BertiCardosoDataFiles}. 
    The last column gives the frequency for $a=0$ with ``naive'' rescaling. 
    We do not show the QNR damping times because extracting those values accurately 
    is difficult and they don't vary much for the moderate range in $a$ that we
    are considering here. }\label{tab:QNRscaling}
\vspace{-.2in}
$$
\begin{array}{|c|l|l|l|l|}\hline
  {q} &{\rm SXS}\ \chi_{\rm rem}&{\rm SXS\ ID} &  \omega_{\rm R}^{\rm Kerr}&(1\!\!+\!\!1/q)\,\omega_{\rm R}^{rm Schw}\\ \hline
  1.5 & 0.6641&  {\rm BBH:007} &0.5175 &0.6228  \\
   2 & 0.6234&  {\rm BBH:0169} &0.5021 &0.5605  \\
  2.5 & 0.5807&  {\rm BBH:0259} &0.4877 &0.5231  \\
  3 & 0.5406&  {\rm BBH:0030} &0.4755 &0.4982  \\
  4 & 0.4716&  {\rm BBH:0182} &0.4567 &0.4671  \\
    5 & 0.4166&  {\rm BBH:0054} &0.4436 &0.4484  \\
  6 & 0.3725&  {\rm BBH:0181} &0.4438 &0.4360  \\
  8 & 0.3067& {\rm BBH:0063} &0.4208 & 0.4204 \\
  9.5 & 0.2708&  {\rm BBH:0302} &0.4142 &0.4130  \\
  \infty & 0& - &0.37367 &0.37367  \\
 \hline
\end{array}
$$
\vspace{-.0in}
\end{table}

Although this particular aspect of the PP/NR comparison is not very useful, it
turns out that a PP-related calculation gives a remarkably accurate estimate of 
the remnant spin. The underlying principle~\cite{BlandfordHughes,Kennefick,BuoKidderLehner} 
is that negligible angular momentum is radiated after the binary enters the plunge. 
For a merger of non-spinning holes, therefore, the angular momentum of the remnant 
should be the orbital angular momentum of the binary at the innermost stable circular 
orbit.

The predicted spin of the remnant is given in Refs.~\onlinecite{BlandfordHughes,Kennefick,BuoKidderLehner} 
but, for completeness, are repeated here for a particle of mass $\mu$, in a circular 
orbit of radius $r$, around a Kerr hole of parameters $M,a$. From Ref.~\onlinecite{BPT} 
the angular momentum $L_z$ is found 
\begin{equation}\label{LzBy_mu}
  \frac{L_z}{\mu}=\frac{M^{1/2}\left(r^2-2a\sqrt{Mr}+a^2\right)\;}
       {r^{3/4}\sqrt{r^{3/2}-3Mr^{1/2}+2aM^{1/2}        \;}}
\end{equation}
and the radius of the innermost stable circular orbit is given by
\begin{eqnarray}
  Z_1&=&1\!+\!(1\!-\!\frac{a^2}{M^2})^{1/3} \left[(1\!+\!\frac{a}{M})^{1/3}\!+\!(1\!-\!\frac{a}{M})^{1/3}\right]\\
  Z_2&=&\sqrt{3a^2/M^2+Z_1^2\;}\\
  r_{\rm ISCO}&=&M\left[3+Z_2-\sgn{a}\sqrt{(3-Z_1)(3+Z_1+2Z_2)\;}\right]\label{risco}.
\end{eqnarray}

In Eqs.~\eqref{LzBy_mu}-\eqref{risco}, ``$M$'' is the total spacetime mass. We use the notation	$[L_z/\mu]$
to mean the expressions above with $M=1$. In our notation the expression $
\left[{L_z}/{\mu
  }\right]
$  is dimensionless and has the meaning                                                                                            
                                                                                                                                
\begin{equation}                                                                                                                
  \left[\frac{L_z}{\mu}\right]\equiv\frac{\mbox{angular momentum}}{(\mbox{particle mass})(M_1+M_2)}                             
\end{equation}                                                                                                                  
We take the particle mass to be  the reduced mass of the system,                                                                
\begin{equation}                                                                                                                
  \mu\equiv \mbox{particle mass}=\mbox{reduced mass}=\frac{M_1M_2}{M_1+M_2}\,,                                                  
\end{equation}                                                                                                                  
In our comparisions we will be using the
dimensionless spin $\chi_{\rm rem}$ of the remnant defined in the SXS catalog as $L_z$ divided by the square
of the total system mass. The comparable quantity in our PP work is the	dimensionless spin
\begin{eqnarray}
  a_{PP}=&\frac{L_z}{(M_1+M_2)^2}&=\left[\frac{L_z}{\mu}\right] \frac{\mbox{particle mass}}{(M_1+M_2)} \nonumber\\
&=& \left[\frac{L_z}{\mu}\right] \frac{M_1M_2}{(M_1+M_2)^2}\nonumber\\
&=&   \left[\frac{L_z}{\mu}\right] \frac{q}{(1+q)^2}\label{aPP}\,.
\end{eqnarray}
 In our table below we
find	this value for given system parameters by computing $[L_z/\mu]$ from   Eqs.~\eqref{LzBy_mu}-\eqref{risco}  ,with $M=1$, and a moderate, but arbitrary, choice for $a$. 
The resulting $a_{PP}$ is then used as the starting point in Eqs.~\eqref{LzBy_mu}-\eqref{risco}, and the procedure is iterated until a stable PP ``prediction" is found for $a_{PP}$.
This result is then compared with the value of $\chi_{\rm rem}$ for the same model.

The accuracy of this approximation can be seen in the comparison with the NR results in the SXS catalog.
\begin{table}[h]
  \caption{Comparison of the Kerr spin parameter for a remnant from NR and PP.  
    For each value of $q$ in the first column, the second column gives the 
    remnant value of the spin parameter $a$ listed in the SXS catalog~\cite{SXScatalog} 
    for an SXS model, given in the third column, with negligible initial spins.
    The last column gives the comparable spin parameter computed by the procedure described in Eqs.~\eqref{LzBy_mu}-\eqref{aPP}, and
    show excellent agreement. }
\vspace{-.2in}
$$
\begin{array}{|c|l|l|l|}\hline
  {q} &{\rm SXS}\ \chi_{\rm rem}&{\rm SXS\ ID} &a_{PP}\\ \hline
  1.5 & 0.6641&  {\rm BBH:007}  &0.6442  \\
   2 & 0.6234&  {\rm BBH:0169}  &0.6092  \\
  2.5 & 0.5807&  {\rm BBH:0259}  &0.5714  \\
  3 & 0.5406&  {\rm BBH:0030}  &0.5350 \\
  4 & 0.4716&  {\rm BBH:0182}  &0.4708  \\
    5 & 0.4166&  {\rm BBH:0054}  &0.4182  \\
  6 & 0.3725&  {\rm BBH:0181}  &0.3753  \\
  8 & 0.3067& {\rm BBH:0063}  & 0.3103 \\
  9.5 & 0.2708&  {\rm BBH:0302}  &0.2742  \\
  \infty & 0& - &0.0   \\
 \hline
\end{array}
$$
\vspace{-.0in}
\end{table}
{\em The excellent agreement of this approximation with NR results supports
the argument~\cite{BlandfordHughes,Kennefick,BuoKidderLehner} that little angular 
momentum is radiated during the plunge and merger.} 

It is interesting to consider how the accuracy of our PP approximation is modified 
if the holes have nonzero initial spin. As an example, we consider the SXS model BBH:0051, 
with $q=3$, with an initially non-spinning smaller hole, but $\chi_1=Lz/(M_1)^2=0.5$ for 
the more massive black hole. We use our procedure in Eqs.~\eqref{LzBy_mu}-\eqref{risco}. 
We add the contribution to $a_{PP}$ due to the angular momentum carried by $M_1$, 
which means adding $\chi_1 M_1^2/(M_1+M_2)^2$. We then use the result as the input for 
the next iteration. The result, the PP prediction, is $0.7056$, in rough agreement with the 
SXS result 0.7551. For SXS model BH:0049, it is the smaller hole that is spinning, with 
$\chi_2=0.5$. In this case, our PP procedure predicts $a_{PP}=0.6217$, in rough agreement 
with the comparable SXS value $0.5773$. 

Ref.~\onlinecite{BuoKidderLehner} gives an extensive exploration of parameter space for 
the remnant spin approximation. Both in that reference and in our two examples above, the accuracy of the simple 
model is worse when there is initial spin rather than no spin. This suggests that tidal interaction during the 
pre-ISCO motion may play a non-negligible role.

\section{Comparison with optimal rescaling of Ref.~\cite{Surr3}}\label{sec:Comp}
In Ref.~\cite{Surr3} an optimization procedure was used to compute the 
rescaling parameters for PP in order to maximize agreement with NR. In 
particular, the PP amplitude was scaled by a parameter $\alpha$ and the 
time (period) by $\beta$. Then $\alpha$ and $\beta$ were computed in order 
to minimize the $L^2$-norm difference between the rescaled PP and NR waveforms. 
As an example, for $q=3$ and for the $h_{22}$ waveform the optimal values of 
$\alpha$ and $\beta$ were both found to be $0.7$ approximately. The $\alpha$ 
and $\beta$ values based on the total mass and barycenter corrections we present 
here are $\alpha = 0.75$ and $\beta = 0.65$. While we do not view a direct 
comparison of these values as meaningful, but the fact that 
the optimized rescaling yields good agreement with NR over the entire 
waveform duration does suggest that there are other (nonlinear) corrections 
that are somehow being addressed through the optimized rescaling procedure 
over the barycenter correction that we have identified here. 

In the context of higher-modes, it has been noted that the value of $\beta$ 
stays the same as the dominant $\ell=m=2$ case, but the value of the amplitude 
rescaling $\alpha$ drops significantly~\cite{Surr3}. For example, for $\ell=m=3$ 
the optimal value of $\alpha=0.4$, a sharp drop from $0.7$ for the $\ell=m=2$ 
case! It has been suggested that the drop in the amplitude rescaling comes from 
the fact that the PP calculation doesn't account for the substantial size of the 
smaller black hole in the context of comparable mass-ratios like $q=3$~\cite{Surr3}. 
In other words, the PP approximation generates artificially higher amplitudes for higher 
modes due to the fact that the PP itself has a wider bandwidth than an object of 
finite size. We argue that while this issue is likely to play a role, it should 
only be significant for much larger values of $\ell, m$ because the wavelength 
of the radiation would need to become comparable to the size of the smaller black 
hole~\cite{Div}. Instead, as we pointed out in Sec.~\ref{sec:barycenter} the barycenter 
corrections we propose here result in a value of $\alpha=0.54$ which is 
strongly suggestive of a large drop from the value of $0.75$ computed for the 
$\ell=m=2$ case.

\section{Conclusions}\label{sec:Conclusions}
As stated in the Introduction section, the comparison of PP results and	NR
results are a resource for probing the interaction of inspiraling
black holes as they draw closer. Exploiting this, however, must	be
done in a way that compares the two methods at physically equivalent
moments of the inspiral. We have pointed out that this has not been
done in the work associated with establishing surrogate
models~\cite{Surr2,Surr3}. The goal of that effort was to use NR results 
as calibration data for high-mass ratio gravitational waveform computations. 
We suggest that such models may be aided by addressing the issues we raise 
above about alignment and physically based rescaling.

Aside from that, possible directions should be considered for future work. 
The loss of spin angular momentum during inspiral, and its connection to
tidal interactions, has already been made in Sec.~\ref{sec:QNR}. Work has already been started in this area~\cite{BuoKidderLehner}. 
We have also pointed out in Sec.~\ref{sec:barycenter}, that the quadrupole 
formula and the equivalent for the octupole are less accurate than might 
be expected. These formulas are slow-motion approximations and are being 
applied in what would appear to be an appropriate setting. Understanding 
why the formulas miss the mark may be very useful in a variety of
applications.

As a much more challenging extension of	the PP/NR comparison, one
might consider binary inspirals of holes with spin angular momentum
not aligned with orbital angular momentum. Such a comparison would be
helpful in understanding how to use a spinning particle as a proxy for
a rotating black hole.

We also expect that there are other	insights that can be extracted from an
extended PP/NR comparison; insights beyond our current imagination.

\medskip

{\em Acknowledgments:}
Thanks to Saul Teukolsky for discussions of NR. We thank Harald Pfeiffer 
for helpful comments and for bringing several references to our attention. 
Thanks go to Tousif Islam and Scott Field for contributions to auxiliary 
computations and for feedback. Special thanks go to Scott Hughes for many 
important discussions and for bringing to our attention Refs.~\cite{BlandfordHughes, Kennefick} 
and other issues.  G.K. acknowledges support from NSF Grants No. PHY-2106755 
and No. DMS-1912716.  



\begin{thebibliography}{11}

\bibitem{PP}
Pranesh A. Sundararajan, Gaurav Khanna, and Scott A. Hughes, Phys. Rev. D 81, 104009 (2010); 
Anil Zenginoglu and Gaurav Khanna, Phys. Rev. X 1, 021017 (2011); Justin McKennon, Gary Forrester, 
and Gaurav Khanna, Proceedings of the NSF XSEDE12 Conference, Chicago (2012); Scott E. Field, 
Sigal Gottlieb, Zachary J. Grant, Leah F. Isherwood, Gaurav Khanna, Commun. Appl. Math. Comput.  
\url{https://doi.org/10.1007/s42967-021-00129-2} (2021).

\bibitem{Harms}
Enno Harms, Sebastiano Bernuzzi, Alessandro Nagar, Anil Zenginoglu, Class. Quantum Grav. {\bf 31}, 245004 (2014).

\bibitem{SF1}
Adam Pound, ``Motion of Small Objects in Curved Spacetimes: An Introduction to Gravitational Self-Force'' 
In: D. Puetzfeld,  C. Laemmerzahl, B. Schutz, (eds) {\em Equations of Motion in Relativistic Gravity}, Fundamental 
Theories of Physics, {\bf 179}, Springer, Cham (2015).
\bibitem{SF2}
Adam Pound, Phys. Rev. D {\bf 95}, 104056 (2017).
\bibitem{SF3}
Leor Barack, Adam Pound, Rep. Prog. Phys. {\bf 82}, 016904 (2019). 
\bibitem{SF4}
Barry Wardell, Adam Pound, Niels Warburton, Jeremy Miller, Leanne Durkan, Alexandre Le Tiec, arXiv:2112.12265 [gr-qc].
\bibitem{SF5}
Adam Pound, Barry Wardell, Niels Warburton, Jeremy Miller, Phys. Rev. Lett. {\bf 124}, 021101 (2020).
\bibitem{SF6}
Niels Warburton, Adam Pound, Barry Wardell, Jeremy Miller, Leanne Durkan, Phys. Rev. Lett. {\bf 127}, 151102 (2021).

\bibitem{EOB1}
Alessandra Buonanno, Thibault Damour, Phys. Rev. D {\bf 59}, 084006 (1999).
\bibitem{EOB2}
Alessandra Buonanno, Thibault Damour, Phys. Rev. D {\bf 62}, 064015 (2000).
\bibitem{EOB3}
Thibault Damour, Piotr Jaranowski, Gerhard Schaefer, Phys. Rev. D {\bf 62}, 084011 (2000).
\bibitem{EOB4}
Thibault Damour, Phys. Rev. D {\bf 64}, 124013 (2001).
\bibitem{EOB5}
Thibault Damour, Piotr Jaranowski, Gerhard Schaefer, Phys. Rev. D {\bf 91}, 084024 (2015).
\bibitem{EOB6}
Alessandro Nagar, James Healy, Carlos O. Lousto, Sebastiano Bernuzzi, Angelica Albertini, Phys. Rev. D {\bf 105}, 124061 (2022).
\bibitem{EOB7}    
Angelica Albertini, Alessandro Nagar, Adam Pound, Niels Warburton, Barry Wardell, Leanne Durkan, Jeremy Miller, arXiv:2208.01049 [gr-qc].


\bibitem{Surr1}
Vijay Varma, Scott E. Field, Mark A. Scheel, Jonathan
Blackman, Davide Gerosa, Leo C. Stein, Lawrence E. Kidder, and 
Phys.~Rev.~Research.{\bf1}, 033015 (2019).

\bibitem{Surr2}
 Nur E. M. Rifat, Scott E. Field, Gaurav Khanna, and
 Vijay Varma,
 Phys. Rev. D {\bf 101}, 081502 (2020).
\bibitem{Surr3}
Tousif Islam, Scott E.~Field, Scott A.~Hughes,
Gaurav Khanna, Vijay Varma, Matthew Giesler, Mark A.~Scheel, Lawrence E.~Kidder, and Harald P.~Pfeiffer,
	arXiv:2204.01972 [gr-qc]. 



\bibitem{hybrid} 
Vijay Varma, Scott E. Field, Mark A. Scheel, Jonathan Blackman, Lawrence E. Kidder, Harald P. Pfeiffer,
Phys. Rev. D {\bf 99}, 064045 (2019). 




  
\bibitem{SXSpaper}
Michael Boyle {\it et al.}, 
Class.~Quant.~Grav. {\bf 36}, 195006 (2019); Abdul Mrue {\it et al.} Phys. Rev. Lett. {\bf 111}, 241104 (2013).

\bibitem{SXScatalog}
SXS Collaboration, 
SXS Waveform Catalog \url{https://data.black-holes.org/waveforms/catalog.html}



\bibitem{BPT}James M.~Bardeen, William H.~Press, and Saul A. Teukolsky,
  Astrophys.~J.~{\bf 178}, 347 (1972).

  


  


\bibitem{RevModPhys} Kip S. Thorne, Reviews of Modern Physics, {\bf52}, No.\,2, Part l, April (l980).

\bibitem{BlandfordHughes}
Scott~A.~Hughes, Robert~D.~Blandford, 
Astrophys. J. {\bf 585}, L101 (2003).

\bibitem{Kennefick}
Kostas Glampedakis, Scott A. Hughes, Daniel Kennefick,
Phys. Rev. D {\bf 66} 064005 (2002). 
        
\bibitem{BuoKidderLehner} 
Alessandra Buonanno, Larry Kidder, and Luis Lehner, Phys. Rev. D {\bf77} 026004 (2008).

\bibitem{DetweilerPoisson}
  Stephen~Detweiler, Eric~Poisson,
Phys. Rev. D {\bf 69}, 084019 (2004).


\bibitem{BertiEtAl}
Emanuele Berti, Vitor Cardoso, and Andrei~O.~Starinets,
Class. Quantum Grav. {\bf 26}, 163001 (2009). 
  
  




\bibitem{MTW36.20}
Charles~W. Misner, Kip~S. Thorne, and John~A. Wheeler, {\em Gravitation} (W. H.
  Freeman, San Francisco, 1973) Eq.~36.20.


\bibitem{BertiCardosoDataFiles}
Emanuele Berti, Ringdown Data \url{https://pages.jh.edu/eberti2/ringdown/}

\bibitem{Div} 
Enrico Barausse, Emanuele Berti, Vitor Cardoso, Scott A. Hughes, Gaurav Khanna, 
Phys. Rev. D 104, 064031 (2021).

\end{thebibliography}
\end{document}